*Strong spin triplet contribution of the first removal state in the insulating regime of*

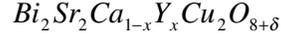

$Bi_2Sr_2Ca_{1-x}Y_xCu_2O_{8+\delta}$

C. Janowitz[1], U. Seidel[1], R.-ST. Unger[1], A. Krapf[1], R. Manzke[1],

V. A. Gavrichkov[2], S. G. Ovchinnikov[2]

1) Humboldt University Berlin, Institute of Physics, 12489 Berlin, Germany

2) L.V. Kirensky Institute of Physics of the Sibirian Branch of the Russian Academy of Science, Krasnoyarsk, 660036, Russia

**Abstract**

The experimental dispersion of the first removal state in the insulating regime of $Bi_2Sr_2Ca_{1-x}Y_xCu_2O_{8+\delta}$ is found to differ significantly from that of other parent materials: oxyclorides and $La_2CuO_4$. For *Y*-contents of $0.92 \geq x \geq 0.55$ due to nonstoichiometric effects in the *Bi-O* layers, the hole concentration in the $CuO_2$-layers is almost constant and on the contrary the crystal lattice parameters *a,b,c* change very strongly. This (*a,b*) parameter increase and *c* parameter decrease results in an unconventional three peak structure at $(0,0); (\pi/2, \pi/2); (\pi, \pi)$ for $x = 0.92$. We can describe the experimental data only beyond the framework of the 3-band *pd*-model involving the representations of a new triplet counterpart for the Zhang-Rice singlet state.

Until now in high-$Tc$ cuprates there was not any clear evidence that there is a contribution to the first removal state from states distinct from the Zhang-Rice singlet (*ZRS*) $A_{1g}$ state. Early theoretical works [1-4] indicating the possible approach of the *ZRS* and the $^3B_{1g}$ two-hole states remained without any experimental support. Interestingly the simple inversion of the triplet and singlet states should be accompanied by a change of the type of the magnetic ordering already in undoped parent structures. The question on the type of magnetic ordering or magnetic and quasiparticle excitations spectra in case of their approach was never investigated since this problem can not be studied within a framework of the 3-band *pd-* and *t-J-* models.

Accordingly theoretical descriptions have been developed for antiferromagnetic (AF) insulators $La_2CuO_4$ and $Sr_2CuO_2Cl_2$ leading to states with the periodicity of the AF Brillouin zone [5,6], i.e. maxima at $\vec{k}=(\pi/2,\pi/2)$. Our systematic high resolution ARPES (angle resolved photoemission) study of the whole insulating range of $Bi_2Sr_2Y_xCa_{1-x}Cu_2O_{8+\delta}$ (x>0.55) shows additional new states at $\vec{k}=(0,0)$ and $\vec{k}=(\pi,\pi)$. In contrast to $La_{2-x}Sr_xCuO_4$, in $Bi_2Sr_2Y_{1-x}Ca_xCu_2O_{8+\delta}$ the hole concentration per Cu $x_h$ is smaller than a substitution concentration x because some holes induced by $Ca^{+2} \rightarrow Y^{+3}$ substitution go to the Bi-O layers. For example in the insulator region $1 \geq x \geq 0.5$ the value $n_h$ changes very weakly by $0.02 \leq n_h \leq 0.05$ [7]. Nevertheless, changes of the crystal lattice parameters are induced by the composition variable x leading to an (a,b)-parameter increase and c-parameter decrease with increasing Y-concentration x [8]. As consequence, the hopping parameters $t_{pp}$ (in plane O - O hopping) and $t'_{pp}$ (in plane O – out of plane O hopping) also vary with the lattice parameters. Thus in $Bi_2Sr_2Y_{1-x}Ca_xCu_2O_{8+\delta}$, at almost constant hole concentration, the reduction of the relation $d_{pl}/d_{ap}$ ($d_{pl}$ - Cu – in plane O and Cu – out of plane distances) or so-called a "chemical" pressure effect takes place. It is, according to [1-4], one of the main reasons for the approach of the singlet $A_{1g}$ and triplet $^3B_{1g}$ states.

$Bi_2Sr_2Y_xCa_{1-x}Cu_2O_{8+\delta}$ single crystals were grown from the melt (for details see [9]). By replacement of the bivalent calcium by trivalent yttrium the hole concentration of the $CuO_2$-planes has been controlled in the samples. The superconducting properties respectively the disappearance of superconductivity with doping were proven by susceptibility measurements. The stoichiometry and, in particular, the Y content was determined by energy dispersive X-ray analysis (EDX). For an Y content of $0.92 \geq x \geq 0.55$ the crystals showed no superconducting transition in susceptibility. The samples were rectangular shaped with the long side along the crystallographic a-axis, confirmed by diffraction experiments, and have a typical size of 5x2 mm$^2$. Crystals were cleaved in UHV (p=1x10$^{-10}$mbar) and were measured at a temperature of 90 K. No effects due to charging of the samples have been observed. LEED and Laue patterns revealed sharp spots for all doping concentrations. An example is given in the panels 1d. and 1e. of FIG. 1. Since all samples showed the about 1x5 reconstruction, the spectra were recorded along the $\Gamma X$ - and not along the $\Gamma Y$ - direction to avoid contributions of superstructure bands. The ARPES-experiments have been performed at the 3m normal-incidence monochromator HONORMI at beamline W3.2 of HASYLAB. For the measurements discussed here 18 eV photon energy was used. The energy distribution curves (EDC) were recorded with a hemispherical deflection analyser with a total acceptance angle of 1° and an energy resolution of 10 meV [10]. Due to the broader emission maxima of insulating HTSC´s an overall resolution ( analyser plus monochromator plus temperature) of 80meV was sufficient in order to improve statistics. An Au Fermi edge served for the Fermi energy reference.

Because the photoemission lineshape of high temperature superconductors at arbitrary doping levels is not well understood there is a strong need for a reliable data analysis procedure that gives a reasonable approximation of the real physics but does not lead to wrong conclusions. It is at present not known whether a spectrum for a given doping level can be interpreted in terms of single-particle excitations. The objective is, therefore, simply to define approximate quantities, which reflect the energy scale of the data. The centroid (the centre of gravity) of spectral features and positions of leading edges are the obvious possibilities. In the recent literature, a rather simple model consisting of a Lorentzian sitting on a step-edge background has been adopted to model the broad dispersing structures observed at the insulating and underdoped cuprates [11]. It has been found out [12] that even in highly underdoped samples the change of the slope of the spectra is a characteristic that easily identifies the broad high energy feature. This procedure is illustrated in FIG 1c. by the dashed intersecting tangents, which approximate the up and down slope of the spectra (tangent method). It is thereby assumed that the ZRS-band is the closer to the Fermi-energy the steeper the slope is. An alternative method is to take the minima of the second derivative of the smoothed spectra (derivative method). This method to determine a dispersion was for instance recently applied by Ronning et al. [5]. An example of this method is given in panel 1b of FIG. 1, which shows a selection of second derivatives from the spectra of panel 1c (the derivatives were multiplied by minus one to obtain peaks). We used this method preferentially. Only in cases, where the second derivatives came out too broad or as double structures, as was the case for the $\Gamma$-spectrum and the three spectra at highest angles from panel 1c., the tangent method was used. A comparison of both methods on the same spectra yielded approximately the same results. The intersection of the two slopes (tangent method) showed a systematic 30-60 meV- shift to higher binding energies when compared to the derivative method. The typical error from the tangent method was 100meV and from the derivative method between 60meV and 100meV. A detailed report will be given in a forthcoming publication.

In FIG. 1 spectra series of the insulating state of the $Bi_2Sr_2Y_xCa_{1-x}Cu_2O_{8+\delta}$ single crystals are shown for the $\Gamma X$-direction $((0,0) \rightarrow (\pi,\pi))$ of the Brillouin zone (panel 1a.). The origin of the dispersing spectral weight near the Fermi level is due to strongly correlated CuO states located in the CuO2-planes. While for low Y-content and optimum doped crystals with highest T$_C$ $(x \leq 0.2)$ the well established Fermi level crossing is observed at about 0.4 $\Gamma X$ (not shown), the insulators with $x \geq 0.55$ investigated here reveal no spectral weight at the Fermi energy. But the dispersing ZRS-band is still present. The centroid of the ZRS-bands of all insulators with Y-content of $0.92 \geq x \geq 0.55$ is shifted to about the same binding energy of 300meV. At half way between $\Gamma$ and $(\pi,\pi)$ these insulators exhibit a distinct maximum in their dispersions, which is most pronounced for x=0.55. With increasing c (decreasing hole concentration) the dispersion curves begin to raise up around the $(0,0)$- and $(\pi,\pi)$- points of the Brillouin zone.

    A dramatic change is observed for the x=0.92 insulator. The centroid of the band has now been shifted to about 600meV and instead of one dominating maximum in the dispersion curve three equally strong maxima are observable at positions $(0,0)$, $(\pi/2, \pi/2)$ and $(\pi,\pi)$ of the Brillouin zone. While the 600meV shift is hard to ascribe to a definite reason and possibly due to pinning by defects [5], the appearance of this new state is exciting and new. Around x=0.92 the insulator is supposed to cross the boundary from the non-AF insulating to the AF insulating phase. In the AF state the next-nearest-neighbor copper atoms have antiparallel spin orientation coupled by a super-exchange interaction via oxygen. The Wigner-Seitz cell then becomes twice as large and, as a consequence, the first Brillouin zone (BZ) is reduced by a factor of two and rotated by 45° [13]. If the underlying AF- Brillouin zone would be the only reason for the change in dispersion, the maximum at $(\pi/2,\pi/2)$ as in the oxychlorides could be explained, but not the developing maxima at $(0,0)$ and $(\pi,\pi)$. The above findings are therefore not similar to the observations in the oxychloride $Sr_2CuO_2Cl_2$

[5,6,14] which has been thought to behave like the parent compound of the high-$T_c$ cuprates. Despite the fact that the absolute maximum of the dispersion curve is also in $Sr_2CuO_2Cl_2$ at 50%$\Gamma X$ the band energy as well as the band width are at variance with the insulating $Bi_2Sr_2Y_xCa_{1-x}Cu_2O_{8+\delta}$ samples.

All our attempts to obtain the 3 peaks in the dispersion at $(0,0)$, $(\pi/2,\pi/2)$, and $(\pi,\pi)$ in the framework of the $t-t'-t''-J$ model failed. That is why we started with a more general model, the 5 band $p$-$d$ model that takes into account $Cu$ $d_{x^2-y^2}$, $d_{z^2}$, in-plane $O$ $p_x$, $p_y$, and apical $O$ $p_z$ single hole atomic states. The effect of strong electron correlations is certainly very important in the insulating phase and in a framework of the multiband $pd$-model the $GTB$ –method takes into account different intra- atomic Coulomb and Hund exchange interactions at Cu and O sites as well as Cu-O nearest neighbour repulsion. While in the 3-band $pd$- model the top of the valence band is formed by a dispersion of holes excited into the ZRS state, the new physics in the multiband $p$-$d$ model results from the $^3B_{1g}$ triplet contribution. The triplet counterpart for the $ZRS$ is also known in the 3 -band $pd$- model with the energy much higher than the $ZRS$, $\Delta E = E_T - E_S \approx 2eV$, so the triplet is not relevant in the low energy region. This irrelevance appears to be a model dependent result. In the multiband model presented here $\Delta E$ sharply decreases due to Hund exchange contributions from $d^\uparrow_{x^2-y^2} d^\uparrow_{z^2}$ configuration and additional bonding with apical oxygen induced $t'_{pd}$ and $t'_{pp}$ hopping ( here " ´ " refers to the apical $p_z$ orbital). For realistic parameters fitted well the $ARPES$ – results for $Sr_2CuO_2Cl_2$ [15] the value $\Delta E \approx 0.7$ eV and excitation of the extra hole added to the $b_{1g}$ initial state to the triplet $^3B_{1g}$ state gives strong admixture near the $(0,0)$ and $(\pi,\pi)$ points to the $ZR$- singlet. To describe the $ARPES$ in the insulating phase of $Bi_2Sr_2Ca_{1-x}Y_xCu_2O_{8+\delta}$ we take into account the strong lattice parameter dependence on the $Y$

– content: parameter c decreases and in-plane parameters $a,b$ increase with the Y-concentration x [8], and neglect the small changes in the hole concentration. The corresponding changes of the in-plane oxygen $t_{pp}$ hopping and the in-plane apical oxygen hopping $t'_{pp}$ are given in the Table 1. For simplicity the other model parameters are the same as in the undoped $CuO_2$ layer [15]. The dispersion of the top of the valence band for different Y concentration has then been calculated by the *GTB* method and is shown in FIG.2. With increasing Y content the three peak structure along the $(0,0) \to (\pi,\pi)$ direction is clearly observed with the $(\pi/2,\pi/2)$ peak slightly decreasing its energy. Along the $(\pi,\pi) \to (\pi,0)$ line there is no significant effect of the Y substitution. These results are in a good agreement to the *ARPES* data. Both states well mix to the one band of first removal states, in spite of the fact that there is essential difference between them.

To clarify the triplet vs. singlet contribution we have calculated partial spectral weight contributions to the *ARPES* peaks (FIG.3). The spectral function for the $(\pi/2,\pi/2)$ peak is determined mostly by the singlet, like in the $t-J$ model. The main contribution to the $(0,0)$ and $(\pi,\pi)$ peaks is given by the triplet $^3B_{1g}$ and this contribution grows with increasing x. FIG. 4 gives a comparison between experiment and theory for the dispersion of the crystals with the highest (x = 0.92) and lowest (x = 0.55) Y- concentration along (0,0) → (π,π). The spectrum for the x=0.92 sample has been shifted to equal minimum binding energy with the x=0.55– spectrum in the manner also applied by Ronning et al. [5]. It can be seen very clearly from FIG. 3 and FIG. 4 that the dispersion at (0,0) and (π,π) changes considerably, because the new contribution of triplet states becomes detectable due to its increased spectral weight.

To conclude, we have measured that due to the "chemical" pressure effect induced by Y-substitution in $Bi_2Sr_2Ca_{1-x}Y_xCu_2O_{8+\delta}$, the dispersion of the first removal state shows, at least near the AF phase at $x=0.92$, a pronounced three peak structure at the $(0,0)$,

$(\pi/2, \pi/2)$, $(\pi, \pi)$ symmetric points of the BZ. Modelling the changes of the *a,b,c* lattice parameters in the *GTB* method with an essential *3-dimensional* 5-band *pd*-model we reproduced the experimental three peaks structure and its concentration dependence. One may say our results indicate that the $(0,0)$ and $(\pi, \pi)$ peaks result from the two-hole $^3B_{1g}$ counterpart for the Zhang-Rice state near $E_F$, which appears at far higher binding energies in *2-dimensional 3*-band *pd*-models or *t-J* models.

Our data also support the earlier scenario [16-17] that the dispersion along the $(\pi/2, \pi/2) \leftrightarrow (\pi, 0)$ direction is strongly reduced by the inclusion of the apical oxygen orbital, and their inclusion is absolutely essential for obtaining the weak dispersion observed experimentally. Thus we offer new good testing ground for the theory of band structure in high-*Tc* cuprates.

**Acknowledgement**

The authors are thankful to Dr. D.Manske and Dr. I.Eremin for their hospitality during the visit to Berlin, Dr. S. Rogaschewski and Dr. H. Dwelk for the characterization of the crystals, the staff of HASYLAB, especially Dr. P. Gürtler and the group of Prof. M. Skibowski from the University of Kiel for assistance with the measurements. This work was supported by INTAS grant 01-0654, RFFI-KKFN "Enisey" grant 02-02-97705, RFFI grant 03-02-16124, Physical Branch of the Russian Academy of Science Program "Strongly correlated electron systems" and by the BMBF (project no. 05 KS1KH11).

*Tables*

| Y-content x | 0 | 0.55 | 0.72 | 0.81 | 0.92 |
| --- | --- | --- | --- | --- | --- |
| $t_{pp}(x)$ | 0.46 | 0.35 | 0.34 | 0.33 | 0.32 |
| $t'_{pp}(x)$ | 0.42 | 0.44 | 0.45 | 0.47 | 0.48 |

Table 1:

$t_{pp}(x)$ and $t'_{pp}(x)$ hopping parameters as used in our calculations, depending on the Yttrium-concentration x.

*Figure captions.*

Fig.1.

1a: EDC's of $Bi_2Sr_2Ca_{1-x}Y_xCu_2O_{8+\delta}$ single crystals in the insulating phase taken along the ΓX direction of the Brillouin zone for different Y- content at T =90 K. With increasing Y concentration the number of holes in the $CuO_2$- plane decreases. The polarization plane of the synchrotron radiation was in the ΓM $((0,0)\rightarrow(\pi,0))$ direction.

1b: Second derivative of a selection of spectra from panel 1c) of FIG. 1, multiplied by a factor of minus one. The maxima are marked by a dot.

1c: Spectra along the $\Gamma \rightarrow X$ $((0,0)\rightarrow(\pi,\pi))$ direction of the sample with the highest Y-concentration (x=0.92). The positions of the centroids have either been obtained from the maxima of panel 1b (derivative method) or as intersection of the tangents (tangent method). In the latter case this has been indicated by two straight lines in panel 1c.

1d: Example of a typical LEED-picture for a single crystal with Y- concentration x=0.72 at an electron energy of 70 eV.

1e: Example of a typical Laue- pattern for a single crystal with Y- concentration x=0.72

1f: Dispersion of the uppermost CuO derived states as obtained from the spectra of panel 1a along the major symmetry lines. The ΓM- dispersions are from spectra not shown.

Fig. 2

The dispersion of the top of the valence band calculated by the *GTB* method for different Y-contents x. Y(x) as indicated in the figure. The respective $t_{pp}(x)$ and $t'_{pp}(x)$ hopping parameters used for the calculations are given in Table 1.

Fig.3

Partial weights of the triplet states (dotted line) and singlet states (solid line) to $A_{tot}$ - total spectral intensity- at two different Y-contents, $x_1 \geq x_4$. Here the spectral function $A(\vec{k},\omega)$ is taken along the peak positions in the $(\vec{k},\omega)$ plane according to the dispersion shown in Fig. 1.

Fig. 4

Comparison of experimental (dots: x=0.55, squares: x=0.92) versus theoretical (green line: $x = x_4 = 0.55$, red line $x = x_1 = 0.92$ from Fig. 2) dispersions for the samples with the highest (x=0.92) and lowest (x=0.55) Y- concentration along $\Gamma \rightarrow X$ $((0,0) \rightarrow (\pi,\pi))$. The experimental x=0.92- dispersion has been shifted to obtain a common valence band maximum with the x=0.55-dispersion. (see also the text).

Fig.1

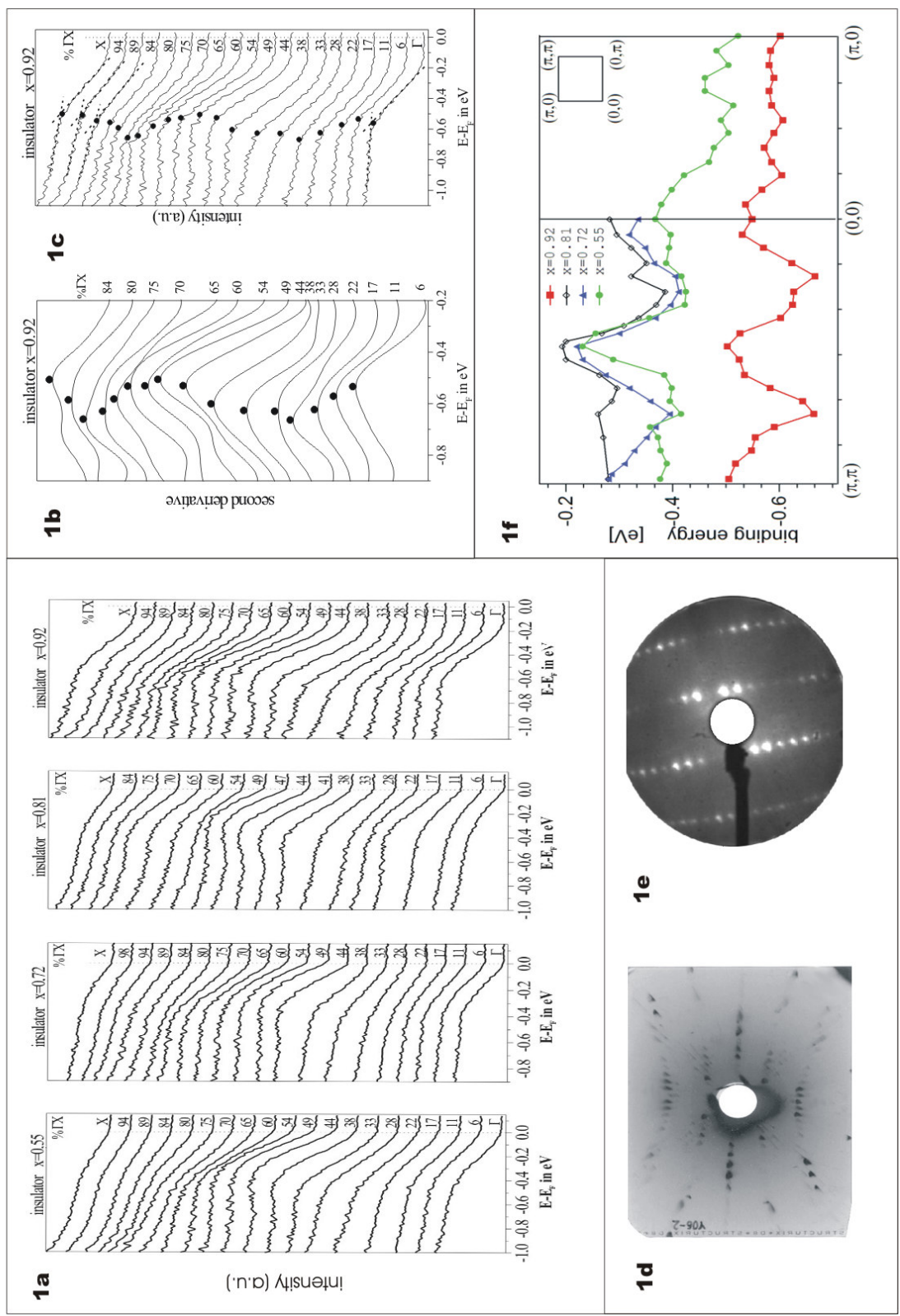

FIG. 2

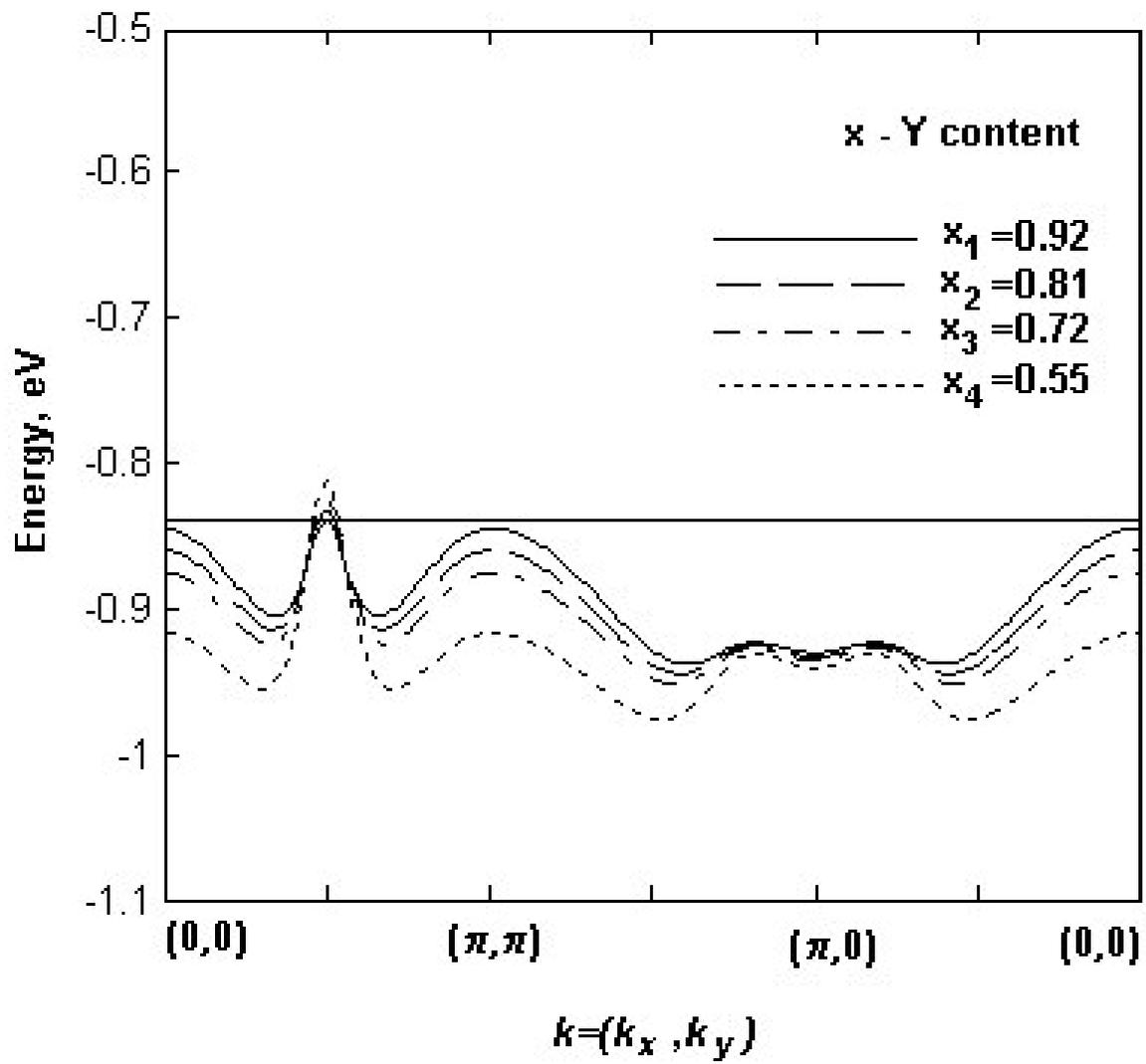

Fig. 3

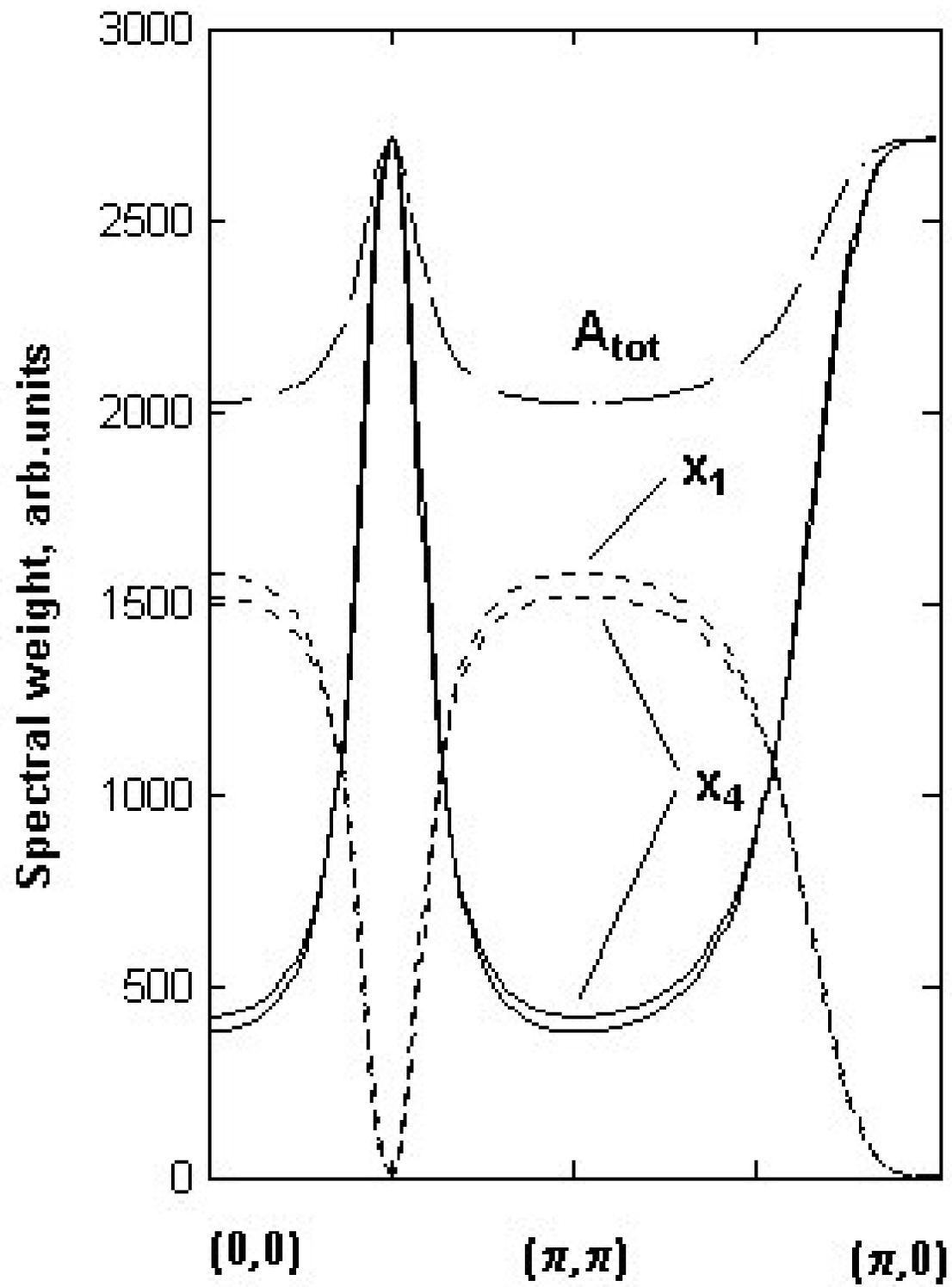

FIG. 4

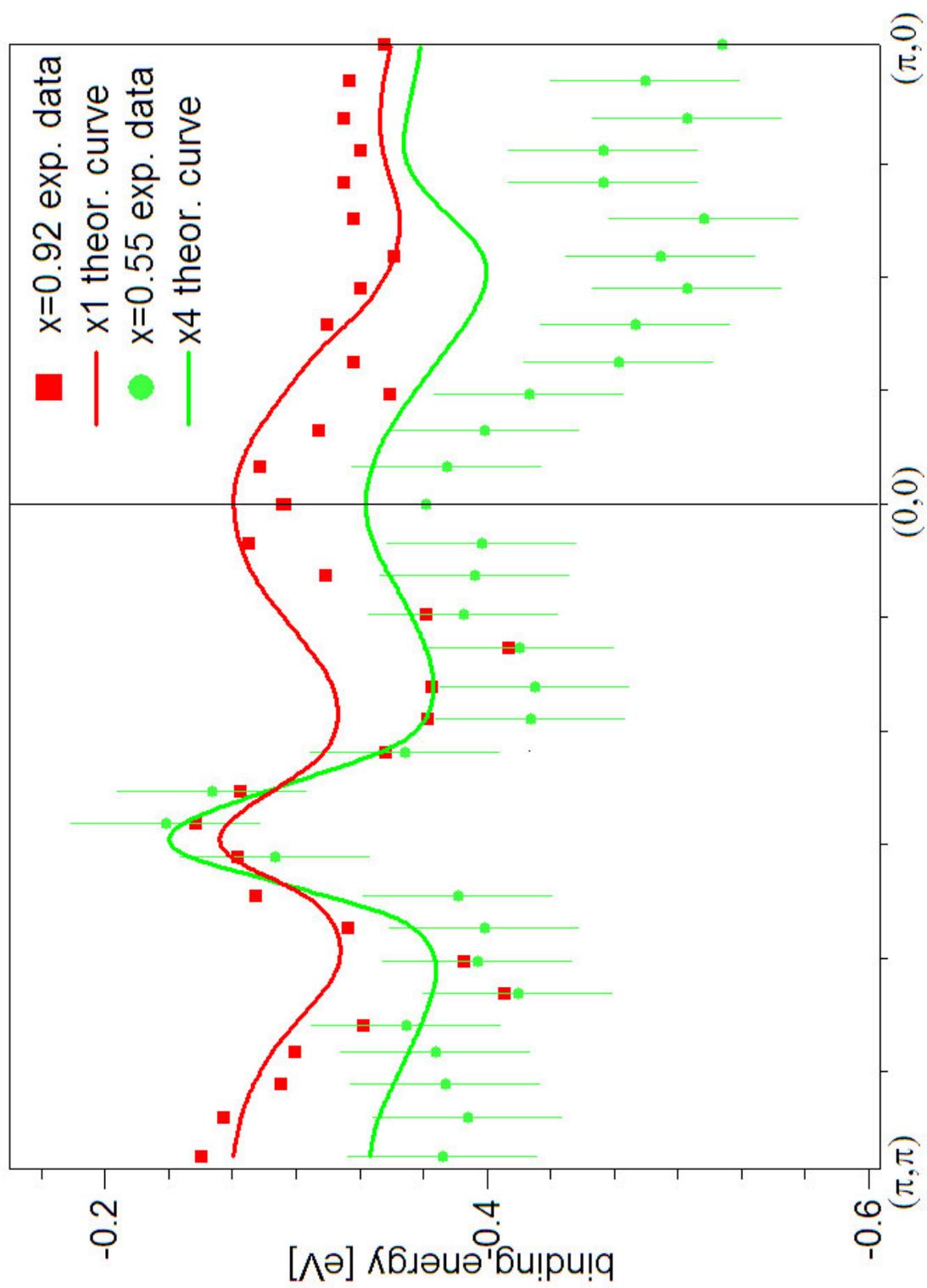